\documentclass[prl,superscriptaddress,preprint,showpacs,aps]{revtex4}

\usepackage{graphicx}
\usepackage{dcolumn}
\usepackage{amsmath}

\begin{document}

\title{Two energy scales and slow crossover in YbAl$_{3}$}
\author{J. M. Lawrence}
\email{jmlawrenc@uci.edu}
\affiliation{University of California, Irvine, CA 92697}
\author{T. Ebihara}
\affiliation{Shizuoka University, Shizuoka 422-8529, Japan}
\author{P. S. Riseborough}
\affiliation{Temple University, Philadelphia, PA 19122}
\author{C. H. Booth}
\affiliation{Lawrence Berkeley National Laboratory, Berkeley, CA 94720}
\author{M. F. Hundley}
\author{P. G. Pagliuso}
\author{J. L. Sarrao}
\author{J. D. Thompson}
\affiliation{Los Alamos National Laboratory, Los Alamos, NM 87545}
\author{M. H. Jung}
\author{A. H. Lacerda}
\affiliation{National High Magnetic Field Laboratory, Los Alamos National
Laboratory, Los Alamos NM 87545}
\author{G. H. Kwei}
\affiliation{Lawrence Livermore National Laboratory, Livermore CA 94551 }

\date{\today}

\begin{abstract}
Experimental results for the susceptibility, specific heat, 4f
occupation number, Hall effect and magnetoresistance for single
crystals of YbAl$_{3}$ show that, in addition
to the Kondo energy scale $k_{B}T_{K}$ $%
\sim $ 670K, there is a low temperature scale $T_{coh}<50$K for
the onset of coherence. Furthermore the crossover from the low
temperature Fermi liquid regime to the high temperature local
moment regime is slower than predicted by the Anderson impurity
model. These effects may reflect the behavior of the Anderson
Lattice in the limit of low conduction electron density.
\end{abstract}

\pacs{75.30.Mb 75.20.Hr 71.27.+a 71.28.+d 61.10.Ht }

\maketitle

The low temperature transport behavior of periodic intermediate
valent (IV) and heavy fermion (HF) compounds\cite{Hewson} is
fundamentally different from that expected for the Anderson
impurity model (AIM), in that it manifests vanishing resistivity
(Bloch's law) and an optical conductivity\cite{Sievers}
appropriate for renormalized band behavior. Similarly, the 4f
electrons have a coherent effect on the Fermi surface, as seen in
de Haas van Alphen (dHvA) measurements\cite{dHvA}. However the
temperature dependence of the susceptibility, specific heat and
4f occupation number and the energy dependence of the dynamic
susceptibility show behavior that is qualitatively very similar
to the predictions of the AIM\cite{BCW,crossover}. Essentially
this is because these properties are dominated by spin/valence
fluctuations that are highly local and which exhibit a Lorentzian
power spectrum\cite{Lawrence} consistent with the AIM.

Nevertheless, recent theoretical studies\cite{Jarrell,Georges} of
the Anderson lattice (AL) in the Kondo limit suggest that these
latter properties can differ in at least two ways from the
predictions of the AIM. First, as the background conduction
electron density $n_{c}$ decreases, the theory
predicts\cite{Georges} a new low temperature scale $T_{coh}$ for
the onset of Fermi liquid coherence, where $T_{coh}$ is
significantly smaller than the high temperature Kondo scale
$T_{K}$. For $T < T_{coh}$ peaks in addition to those expected on
the high temperature scale $T_{K}$ are predicted for the
susceptibility and specific heat. Second, as $n_{c}$ decreases
the crossover from low temperature Fermi liquid behavior to high
temperature local moment behavior becomes slower (i.e. more
gradual) than predicted for the AIM\cite{Jarrell}.

 Recently we have given evidence\cite{crossover} for a slow crossover of the
 susceptibility and 4f
occupation number in the IV compounds Yb$X$Cu$_{4}$ ($X$= Ag, Cd,
Mg, Tl, Zn). In this paper we give evidence {\it both} for a slow
crossover {\it and} for two energy scales in the IV compound
YbAl$_{3}$. We show that both effects are observed not only for
the susceptibility but also for the specific heat.  We show that
the magnetotransport gives clear evidence for a change in
character on the low temperature scale and also suggests that the
conduction electron density is low ($n_{c}\sim~0.5/atom$).

The samples were single crystals of YbAl$_{3}$ and LuAl$_{3}$
grown by the ''self-flux'' method in excess Al. The quality of
such samples, as measured by the resistance ratio
($R(300$K$)/R(2$K$)\sim ~60$) is sufficiently high that dHvA
signals are well-resolved\cite{Ebihara}. The susceptibility was
measured using a SQUID magnetometer and the specific heat was
measured via a relaxation technique. The Hall coefficient was
measured in a magnetic field of 1T using an a.c. resistance
bridge. The magnetoresistance was measured at the Los Alamos
Pulsed Field Facility of the National High Magnetic Field
Laboratory using a 20T superconducting magnet and an a.c. bridge.
The $4f$ occupation number $n_{f}(T)$ was determined from the Yb
$L_{3}$ x-ray absorption near-edge structure, measured at the
Stanford Synchrotron Radiation Laboratory (SSRL) on beam line
4-3; the technique for extracting $n_{f}$ from the data and other
experimental details were similar to those discussed earlier
\cite{Sarrao}. We note here that the Lu $L_{3}$ near-edge
structure measured for LuAl$_{3}$ was used a standard.

\begin{figure}[t]
\includegraphics[width=3.3in]{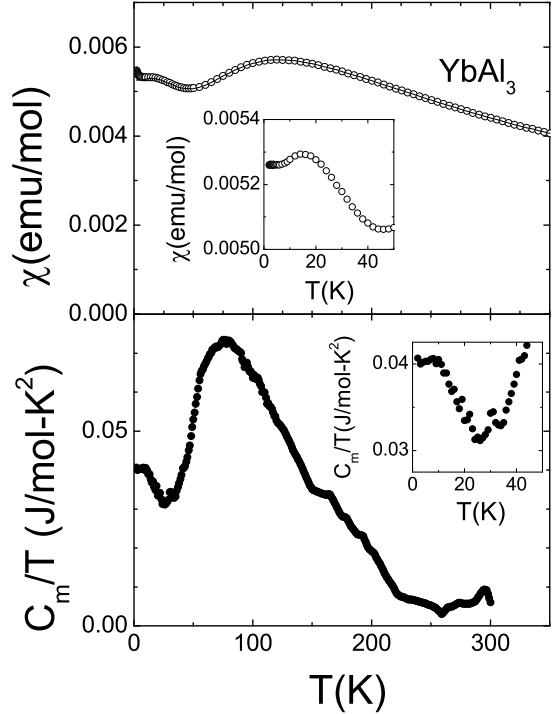}
\caption{ (a) The magnetic susceptibility $\protect\chi (T)$ and
(b) the magnetic contribution to the specific heat coefficient
$C_{m}/T$ for YbAl$_{3}$. The insets exhibit the low temperature
behavior; a small Curie tail has been subtracted from the data in
the inset for $\chi(T)$.}
\end{figure}

In Fig. 1 we plot the susceptibility $\chi(T)$ and the linear
coefficient of the 4f specific heat, where $\gamma _{m}$ =
$C_{m}/T$ and $C_{m}$ = $C$(YbAl$_{3}$) - $C$(LuAl$_{3}$). (For
LuAl$_{3}$ at low temperatures, $C=\gamma T+\beta
T^{2}$ with $\gamma =4$mJ/mol-K$^{2}$ and $\beta =$1.15x10$^{-4}$J/mol-K$%
^{4} $, which implies a Debye temperature $\Theta _{D}=257$K.)
The broad peaks near 100K are typical of Yb IV compounds with
$T_{K}$ $\sim $ 500K. In addition, there is a peak at 15K in the
susceptibility (first reported by Heiss {\it et al.}\cite{Heiss})
and the specific heat coefficient displays an upturn below 25K
which saturates at $T=0$. These additional features are the basic
evidence for the existence of a low temperature scale,
$T_{coh}\lesssim 50$K below which there is a significant change
in the behavior of the compound.

Evidence for this change of character can also be seen in the
magnetotransport (Fig. 2). The Hall coefficient of LuAl$_{3}$ is
typical of a metal, being constant and of a magnitude that in a
one-band model implies a carrier density $n_{c}=1/eR_{H}=
2e^{-}/f.u.$. The high temperature Hall coefficient of YbAl$_{3}$
varies with temperature in a manner suggestive of scattering from
Yb moments (although the data cannot be fit well with the standard
skew scattering formula\cite{Cattaneo}). Near 50K the derivative
$dR_{H}/dT$ of the Hall coefficient changes sign. Similar
anomalies in the low temperature Hall coefficient have been
observed in a number of Ce compounds and have been attributed to
the onset of coherence\cite{Cattaneo}. The magnetoresistance
(Fig. 2b, inset) follows a $B^{2}$ law above 50K and the
magnitude is approximately the same for field parallel and
transverse to the current.
 Below 50K, the magnetoresistance becomes more nearly linear and the transverse
 magnetoresistance becomes substantially larger ($\Delta R /R \sim ~0.75 $) than
 the parallel magnetoresistance ($\Delta R /R \sim ~0.35$) at 2K.  In Fig. 2b we
 plot $\Delta R/R$ versus $Br_{0}$ where $r_{0} = R(150K)/R(T)$; this tests Kohler's
 rule, i.e. that at any temperature $\Delta R/R=A\,f(Br_{0})$ depends only on the
 product $Br_{0}$.  The data
 violate this rule essentially because A varies with T, increasing by a factor of
 almost 1.5 between 40 and 80K; this crossover is seen most clearly in the data
 measured as a function of temperature at a fixed field 17.5T. These magnetotransport
 anomalies suggest that the anomalies in $\chi$ and $C/T$ may
  be associated with an alteration of the Fermi surface.  The fact
  that the resistivity follows a $T^{2}$ law below 30K\cite{Ebihara} suggests that
  the anomalies are also connected with the full establishment of Fermi liquid
  coherence.

To demonstrate that the crossover from Fermi liquid behavior to
local moment behavior is slower than predicted by the AIM we
compare the experimental data to the AIM result that follows from
the measured ground state values of the susceptibility and $4f$
occupation number\cite{Jarrell,crossover}. For fixed spin-orbit
splitting ($\Delta _{so}=$ 1.3eV), the theoretical results
calculated in the non-crossing approximation (NCA) depend on
three parameters: the $f$-level position $E_{f}$, the
hybridization $V$ between the 4$f$ and the conduction electrons,
and the width $W$ of the conduction band, assumed Gaussian
$N(E)=e^{-E^{2}/W^{2}}/\sqrt{\pi }W$. Since only states within
$k_{B}T_{K}$ of the Fermi energy contribute significantly to the
results, we argue that use of a Gaussian bandshape is acceptable
as long as the density of states at the Fermi energy is chosen
appropriately. We choose $W$ to give the same specific heat
coefficient that we observe in
LuAl$_{3}$ ($\gamma $ = 4mJ/mol-K$^{2}$). Once $W$ is fixed, the values of $%
E_{f}$ and $V$ that yield the appropriate $n_{f}(0)$ and $\chi $(0) can be
determined uniquely. The values of the parameters are given in Fig.3; $T_{K}$
is determined from the formula
\[
T_{K}=(\frac{V^{2}}{\sqrt{\pi }W\mid E_{f}\mid })^{1/8}(\frac{W}{\Delta _{so}%
})^{6/8}We^{\sqrt{\pi }WE_{f}/8V^{2}}
\]
which includes the effect of spin orbit splitting but ignores
crystal field splitting since for IV compounds $T_{K}$
\mbox{$>$}%
\mbox{$>$}%
$T_{cf}$. It is clear from Fig. 3 that the susceptibility, $4f$ occupation
number and $4f$ entropy $S_{m}=\int dT$ $C_{m}/T$ all {\it qualitatively}
follow the predictions of the AIM. \ The calculated coefficient of specific
heat ($\gamma =$ 47.8mJ/mol-K$^{2}$) is within 20\% of the measured value ($%
\gamma _{m}=$ 40.65mJ/mol-K$^{2}$). \ Indeed the data even are in
accord with the prediction that the entropy $S_{m}(T)$\
approaches the high temperature limit faster than the effective
moment $T\chi /C_{J}$\ which in turn evolves more rapidly than
$n_{f}(T)$ (see Fig. 3). \ Nevertheless, it is also clear that
the experimental data for these quantities approach the high
temperature limit considerably more slowly than predicted by the
AIM theory.

\begin{figure}[t]
\includegraphics[width=3.3in]{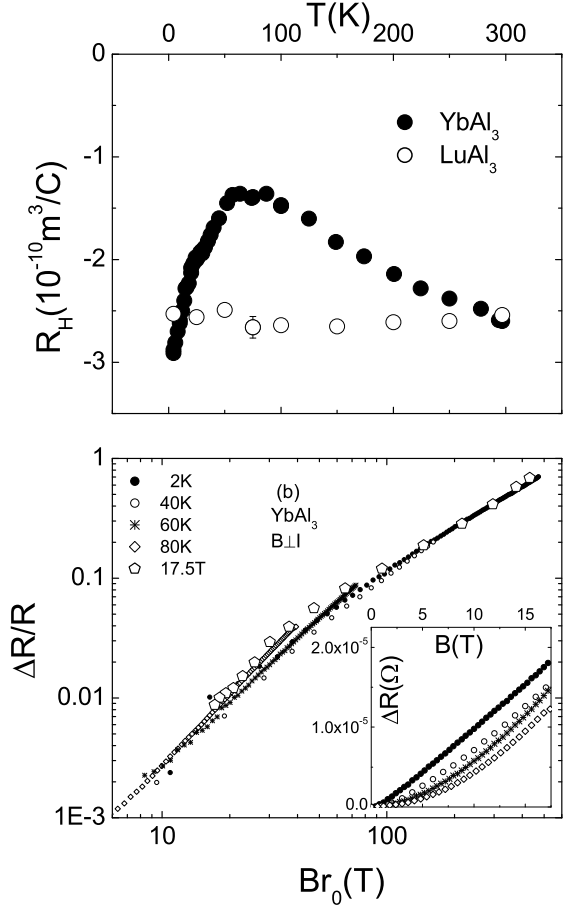}
\caption{ a) The Hall coefficient of YbAl$_{3}$ (closed circles)
and LuAl$_{3}$ (open circles) versus temperature. For LuAl$_{3}$
a typical error bar is shown; for YbAl$_{3}$ the error is smaller
than the size of the symbols. b) Inset: The transverse
magnetoresistance $\Delta R/R=R(H,T)-R(0,T)$ versus magnetic
field for four temperatures. Main panel: The relative
magnetoresistance $\Delta R/R$ versus $Br_{0}$ where
$r_{0}=R(150K/R(T))$.}
\end{figure}

Thus we have demonstrated the existence of a new low temperature
scale for YbAl$_{3}$ and we have shown that the crossover to local
moment behavior is slower than expected based on the Anderson
Impurity Model. In the theory of the Anderson lattice, such
results are expected\cite{Jarrell,Georges} as the conduction
electron density decreases from the value $n_{c}$ = 1 appropriate
to a half-filled band when $n_{f}=1$. In our earlier
work\cite{crossover} on Yb$X$Cu$_{4}$ we assumed that the
background band in the Yb compound is the same as the measured
band in the corresponding Lu compound.  Using a one-band formula
$n_{c}=1/eR_{H}$(Lu$X$Cu$_{4}$) we found that the slow crossover
emerged when the number of electrons per atom (the number per
formula unit divided by the number of atoms in the formula unit)
decreased below the value unity. Strong deviations occurred already for YbMgCu%
$_{4}$ where $n_{c}\sim $ ~0.5/atom. Using the same approximations for YbAl$%
_{3}$, we deduce from the Hall coefficient of LuAl$_{3}$ (Fig. 2a) that $%
n_{c}\sim $ ~ 0.5/atom. Hence, while the conduction electron
density is not low for YbAl$_{3}$, it is (in this approximation)
as low as in other compounds where strong deviations from the AIM
are observed.

We believe that such effects are generic to IV compounds.  A
number of years ago we gave evidence\cite{CePd3} for a small
coherence scale in the IV compound CePd$_{3}$ based on a low
temperature peak in the susceptibility and the extreme
sensitivity of the transport behavior to Kondo hole impurities
below 50K. A low temperature anomaly in the Hall coefficient also
has been observed for this compound\cite{Cattaneo}. Optical
conductivity measurements\cite{Sievers,Bucher,Degiorgi} showed
both that the temperature scale for these effects is the same as
for the renormalization of the effective mass and the onset of the
hybridization gap and that the conduction electron density of
CePd$_{3}$ is less than 0.1 carrier/atom.  Hence the anomalies
appear to be associated both with the onset of the fully
renormalized coherent ground state and with low conduction
electron density. Measurement of the infrared optical
conductivity is clearly a key future experiment for YbAl$_{3}$.

 Neutron scattering measurements on polycrystals of YbAl$_{3}$\cite
{Murani} indicate that in addition to the usual Lorentzian
excitation centered at $E_{0}$ = 40meV, corresponding to the high
temperature Kondo scale (for the parameters of Fig. 3 the AIM predicts $%
E_{0} $ = 39.6meV), a new excitation centered at 30meV arises
below 50K. Preliminary results\cite{Preliminary} in single
crystals also exhibit this excitation. However, the scattering
does not exhibit the expected decrease with increasing $|{\bf
Q}={\bf q}+{\bf G}|$ (where ${\bf G}$ is a reciprocal lattice
vector) for Yb $4f$ scattering, which should follow the Yb 4f
form factor. Indeed the scattering appears to be independent of
${\bf G}$. Further work needs to address this issue.

The theoretical work on the Anderson lattice which predicts slow
crossover and a low temperature coherence scale was motivated by
the desire to understand ''Nozi\'{e}res' exhaustion''. This
concept was raised\cite{Nozieres} to try to understand how the
conduction electrons could screen the $4f$ spins when the number
of conduction electrons $n_{c}$ is smaller than the number
$n_{f}$ of $4f$'s in the lattice. The theoretical work to date
has all been performed in the Kondo limit.  We have examined the
extension of the slave boson mean field theory for the Anderson
lattice to the case $n_{f}<1$ and $n_{c}<1$ relevant to
YbAl$_{3}$, where $n_f$ is the number of holes in the $(2J+1)$
fold degenerate f level. Following Millis and Lee\cite{Millis}, we
define the Kondo temperature $k_B T_K$ as the energy of the
renormalized f level relative to the Fermi-energy. The coherence
scale is defined as the renormalized (quasiparticle) $T=0$
bandwidth which for a background conduction band density of
states $\rho$ is given by $k_B T_{coh}=\rho\tilde{V}^{2}$ where
$\tilde{V}$ is the renormalized hybridization
$\tilde{V}=\sqrt{1-n_{f}}V$. In the limit
$(\rho\tilde{V})^{2}<<n_{c}n_{f}/(2J+1)^{2}$ we find that
$T_{coh}/T_{K}=n_{f}/(2J+1)$ independent of $n_{c}$.  For
YbAl$_{3}$ this means that $T_{coh}$ should be an order of
magnitude smaller than $T_{K}$, in qualitative agreement with the
data.

In any case we assert that the two energy scales and slow
crossover predicted by the theory are features of our data, that
these effects show some correlation with a standard measure of the
conduction electron density and that they may be generic for IV
compounds.

\begin{figure}[t]
\includegraphics[width=3.3in]{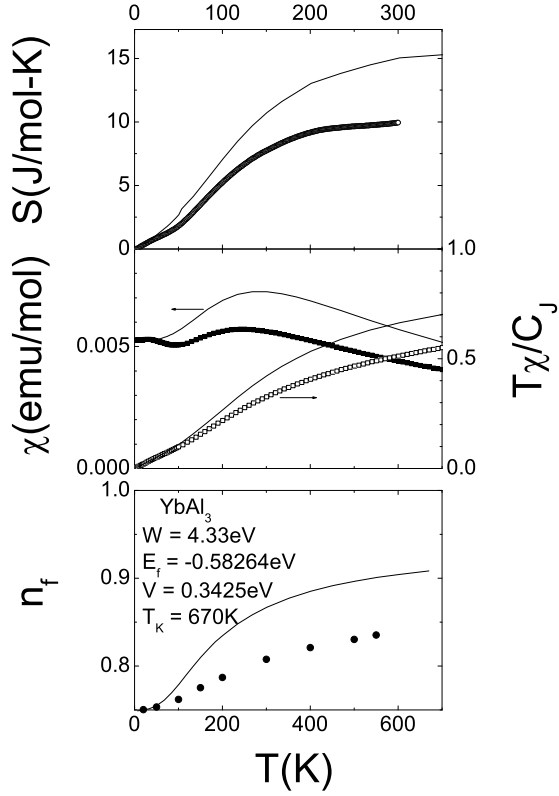}
\caption{ (a) The 4$f$ entropy $S_{m}$, (b) the susceptibility $\protect\chi%
(T)$ (solid symbols) and the effective moment $T\protect\chi /C_{J}$ (open
symbols) where $C_{J}$ is the $J$ = 7/2 Curie constant and (c) the 4$f$
occupation number $n_{f}(T)$ for YbAl$_{3}$. The symbols are the
experimental data and the solid lines are the predictions of the Anderson
Impurity Model (AIM), with input parameters given in the figure. }
\end{figure}

Work at UC Irvine was supported by UCDRD funds provided by the
University of California for the conduct of discretionary
research by the Los Alamos National Laboratory and by the UC/LANL
Personnel Assignment Program. Work at Polytechnic was supported
by DOE FG02ER84-45127. \ Work at Lawrence Berkeley National
Laboratory was supported by DOE Contract No. DE-AC03-76SF00098.
Work at Los Alamos was performed under the auspices of the DOE.
The x-ray absorption experiments were performed at SSRL, which is
operated by the DOE/OBES.


\begin{references}
\bibitem{Hewson}  A. C. Hewson, {\it The Kondo Problem to Heavy Fermions}
(Cambridge University Press, 1993) p. 315.



\bibitem{Sievers}  B. C. Webb, A. J. Sievers and T. Mihalisin, Phys. Rev.
Lett. {\bf 57}, 1951 (1986).

\bibitem{dHvA}  A. Hasegawa and H. Yamagami, Progress of Theoretical Physics
Supplement {\bf 108}, 27 (1992).

\bibitem{BCW}  N. E. Bickers, D. L. Cox and J. W. Wilkins, Phys. Rev. B {\bf %
36}, 2036 (1987).

\bibitem{crossover}  J. M. Lawrence, P. S. Riseborough, C. H. Booth, J. L.
Sarrao, J. D. Thompson and R. Osborn, Phys. Rev. B {\bf 63}, 054427 (2001).

\bibitem{Lawrence}  J. M. Lawrence, S. M. Shapiro, J. L. Sarrao and
Z. Fisk, Phys. Rev. B {\bf 55}, 14467 (1997); J. M. Lawrence, R.
Osborn, J. L. Sarrao and Z. Fisk, Phys. Rev. B {\bf 59}, 1134
(1999).

\bibitem{Jarrell}  A. N. Tahvildar-Zadeh{\it , }M. Jarrell and J. K.
Freericks, Phys. Rev. B {\bf 55}, R3332 (1997); Phys. Rev. Lett. {\bf 80},
(1998) 5168.

\bibitem{Georges}  S. Burdin, A. Georges and D. R. Grempel, Phys. Rev. Lett.
{\bf 85}, 1048 (2000).

\bibitem{Ebihara}  T. Ebihara, S. Uji, C. Terakura, T. Terashima, E.
Yamamoto, Y. Haga, Y. Inada and Y. Onuki, Physica B {\bf 281\&282}, 754
(2000); T. Ebihara, Y. Inada, M.Murakawa, S. Uji, C. Terakura, T. Terashima,
E. Yamamoto, Y. Haga, Y. Onuki and H. Harima, Journ. Phys. Soc. Japan {\bf 69%
}, 895 (2000).

\bibitem{Sarrao}  J. L. Sarrao, C. D. Immer, Z. Fisk, C. H. Booth, E.
Figueroa, J. M. Lawrence, R. Modler, A. L. Cornelius, M. F.
Hundley, G. H. Kwei, J. D. Thompson and F. Bridges, Phys. Rev. B
{\bf 59}, 6855 (1999).

\bibitem{Heiss}  A. Heiss, J. X. Boucherle, F. Givord and P. C. Canfield,
Journal of Alloys and Compounds {\bf 224}, 33 (1995).

\bibitem{Cattaneo}  E. Cattaneo, Z. Physik B {\bf 64}, 305 (1986); {\it ibid}, 317.

\bibitem{CePd3}  J. M. Lawrence, Y.-Y. Chen and J. D. Thompson, in {\it %
Theoretical and Experimental Aspects of Valence Fluctuations and
Heavy Fermions}, edited by L. C. Gupta and S. K. Malik (Plenum
Press, New York, 1987) p. 169.


\bibitem{Bucher}  B. Bucher, Z. Schlesinger, D. Mandrus, Z. Fisk,
J. Sarrao, J. F. DiTusa, C. Oglesby, G. Aeppli and E. Bucher,
Phys. Rev. B \textbf{53}, R2948 (1996).

\bibitem{Degiorgi} L. Degiorgi, F. B. B. Anders and G. Gr\"{u}ner,
Eur. Phys. J. B \textbf{19}, 167 (2001).

\bibitem{Murani}  A. P. Murani, Phys. Rev. B \textbf{50}, 9882 (1994).

\bibitem{Preliminary}  S. M. Shapiro, private communication; E. A.
Goremychkin and R. Osborn, private communication.

\bibitem{Nozieres}  P. Nozi\'{e}res, Ann. Phys. (Paris) {\bf 10}, 19 (1985).

\bibitem{Millis} A. J. Millis and P. A. Lee, Phys. Rev. B \textbf{35},
3394 (1987).

\end{references}
\end{document}